%% file: main.tex
\documentclass[english] {sophia_rcs}

\title{Ab Initio Study of Erbium Point Defects in 4H-SiC for Quantum Devices}

\author{Michael Kuban\\
	\normalsize Carnegie Mellon University\\ Electrical and Computer Engineering\\ Pittsburgh, PA 15213\\
}


\begin{document}

\maketitle

\lhead{}
\chead{}
\rhead{}

\input{sections/abstract}
\input{sections/introduction}
\input{sections/dft}
\input{sections/SiC}
\input{sections/ErInSiC}
\input{sections/conclusion}
\input{sections/acknowledgments}
\input{sections/bibliography}

\end{document}

%% file: sections/abstract.tex
\begin{abstract}
Identifying scalable materials systems that exhibit quantum behavior is a central challenge in quantum information science. Point defects in certain wide-bandgap semiconductors are promising in this regard due to the maturity of semiconductor manufacturing and ion implantation technology. Single erbium defect centers in 4H-SiC are examples of such defects that provide access to discrete defect-induced electron energy levels within the bulk material bandgap, which can be utilized for a variety of quantum technologies, such as single-photon emission for secure communication and distributed quantum computing. This work presents a first-principles study of erbium point defects in 4H-SiC using density functional theory. These results provide materials-level support for the development of Er point defects in 4H-SiC as a scalable platform for quantum devices, helping to bridge the gap between quantum physics and the practical realization of quantum networks.

\end{abstract}

%% file: sections/introduction.tex
\section{Introduction} \label{sec:introduction}

The rapid transition of quantum research from foundational physics to a functional engineering discipline is driven by the unique performance advantages that non-classical behavior can provide over classical behavior. Intentionally harnessing quantum phenomena such as superposition and entanglement to create new devices can lead to transformative technologies in the fields of computing, sensing, and secure networking \cite{future}. However, a primary challenge in realizing quantum devices lies in designing architectures that can be sufficiently isolated to preserve quantum coherence while remaining accessible to reliable control and readout.

One promising modality for isolating a quantum system comes in the form of point defects in solid-state wide-bandgap semiconductors \cite{def}. When a single point defect is created in an otherwise unperturbed bulk crystal lattice, discrete, defect-induced electron energy levels arise that alter the band structure of the material. When the bulk material is a wide-bandgap semiconductor, it becomes possible to have these energy levels arise within the bandgap, providing energetic isolation from electrons in the bulk material. Embedding this defect within a solid-state host also provides the system with mechanical and electrical stability.

The nitrogen-vacancy (NV) defect in diamond is one of the most-studied examples of this, and it has been shown to produce exceptional results for a variety of quantum applications \cite{dia}. However, diamond has some limitations that may prevent it from becoming a scalable platform for quantum technology. One potential issue with this platform is that the emission wavelength of the NV center lies in the visible spectrum ($\lambda\simeq 637$ nm). Although this facilitates visual optical alignment in the laboratory, it integrates poorly with existing large optical telecommunications networks, where emission in the C-band is preferred for achieving low loss in standard optical fibers. C-band emission near 1.55~$\mu$m offers fiber attenuation as low as 0.2~dB/km, compared to losses exceeding 8 dB/km near 637 nm ~\cite{source}. 

Another potential issue with NV centers is the high cost and lack of large-scale processing technology for diamond. In order for quantum devices to be manufactured at scale, it is important that the required materials are available at reasonable cost and compatible with complimentary metal-oxide-semiconductor (CMOS) fabrication processes for integration of required electronic control circuitry.

One possible solution to these issues is to use optically active rare-earth ions embedded within a more common semiconductor host, such as silicon carbide (SiC) \cite{cast}. Unlike diamond, large, high-quality SiC wafers are commercially available at reasonable cost due to their use in power electronics. Additionally, mature ion implantation techniques exist for SiC, allowing for deterministic placement of defects. 

Rare-earth ions are attractive defects for this purpose due to their partially filled 4f shells, which are strongly shielded from the surrounding lattice and provide temperature-invariant atom-like optical transitions \cite{rare}. In particular, erbium ions have been reported to exhibit a zero-phonon line (ZPL) near 1.54 \si{\mu\meter} (0.8 eV) in SiC \cite{emi}. This coincides with the low loss window of standard optical fibers, allowing for ease of integration with existing optical networks \cite{er}.

This work investigates the use of the density functional theory (DFT) method to study erbium-related point defects in the 4H polytype of SiC (4H-SiC) as a potential platform for scalable quantum devices.

%% file: sections/dft.tex
\section{Calculation Specifications} \label{sec:dft}

All calculations in this work were performed using Bridges-2, a high-performance computing node at the Pittsburgh Supercomputing Center \cite{bridge}. The DFT steps were implemented using the ABINIT Software Suite with projector augmented wave (PAW) pseudopotentials \cite{abinit}. For each material structure, a self-consistent field (SCF) calculation was first performed to determine the ground-state charge density at two k-points. Then, the density from the SCF calculation was used in a non-self-consistent (NSCF) calculation to compute eigenvalues across a range of 113 k-points for the band structure and density of states (DOS) plots. The convergence of the calculations was defined as the point when the energy difference between subsequent iterations reached $10^{-4}$ Ry \cite{ErbiumPaper}. A maximum iteration number of 100 was arbitrarily chosen to prevent infinite calculations. This number was only reached for one calculation, which will be discussed further in \cref{sec:ErInSiC}

For pristine 4H-SiC, electronic structure calculations were performed using the Perdew-Burke-Ernzerhof generalized gradient approximation (PBE-GGA) functional \cite{pbe}. Calculations were performed for the 8-atom relaxed unit cell of 4H-SiC, which was obtained from the Materials Project database (mp-11714) \cite{mat}.

For the erbium-related defect calculations, a different approach was necessary. First, in order to approximate a single point defect, a 4x4x1 supercell was used with a single erbium ion placed at the center, replacing a silicon atom. Details of the different defect configurations studied will be discussed in \cref{sec:ErInSiC}

Additionally, due to the presence of strongly correlated and localized 4f electrons in erbium, the GGA+U method was used, where the Hubbard U correction increases the interaction representation of the erbium 4f electrons \cite{hub}. A Hubbard U parameter of 7.21 eV was used for this calculation \cite{ErbiumPaper}.

All band structure and density of states (DOS) calculations were performed along a high-symmetry path through the hexagonal Brillouin zone to visualize the electronic dispersion relations. The path follows the sequence $\Gamma$-M-K-$\Gamma$-A-L-H-A. All plots have been normalized to the valence band maximum (VBM).

%% file: sections/sic.tex
\section{Pristine 4H-SiC} \label{sec:SiC}

Silicon carbide is a wide-bandgap semiconductor with about 250 known polytypes \cite{sic}. 4H-SiC is the preferred polytype for high-power and high-frequency electronics due to its high electron mobility and thermal conductivity. For this reason, 4H is the SiC polytype studied in this paper.

\subsection{Crystal Structure}
\begin{figure}[ht]
\centering
\includegraphics[height=0.55\linewidth,width=0.3\linewidth]{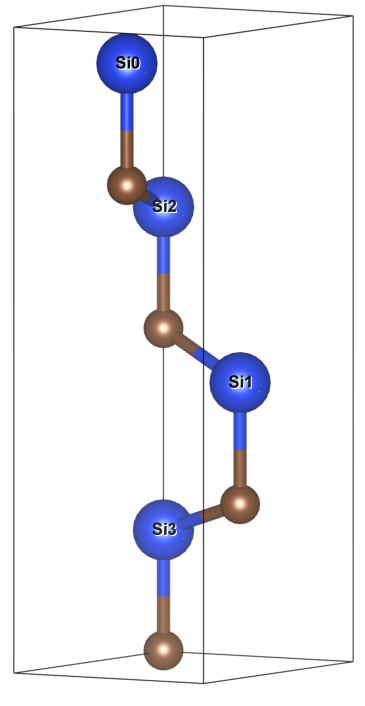}
\includegraphics[height=0.55\linewidth,width=0.5\linewidth]{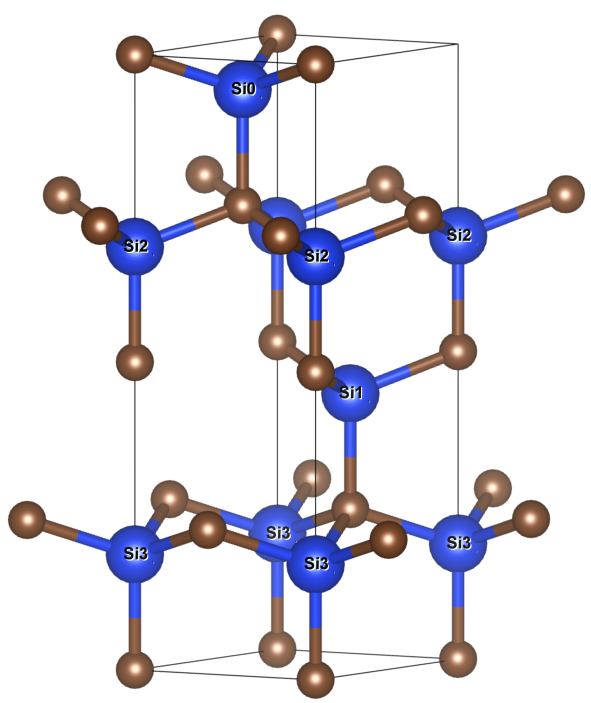}
\caption{4H-SiC unit cell (left) and 4-layer stacking sequence (right).}
\label{fig:unit}
\end{figure}

\begin{figure}[ht]
\centering
\includegraphics[height=0.55\linewidth,width=0.64\linewidth]{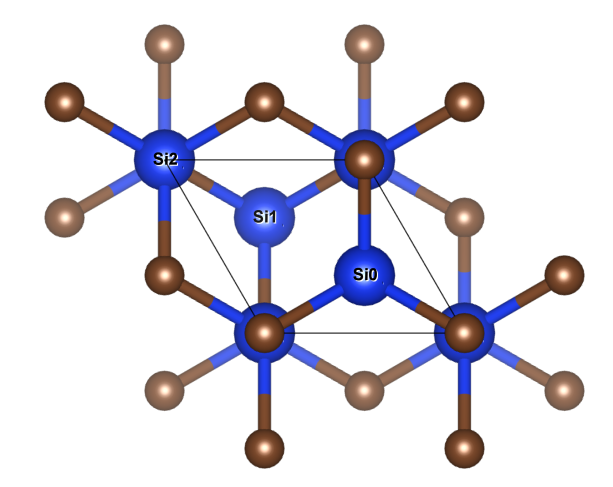}
\includegraphics[height=0.55\linewidth,width=0.35\linewidth]{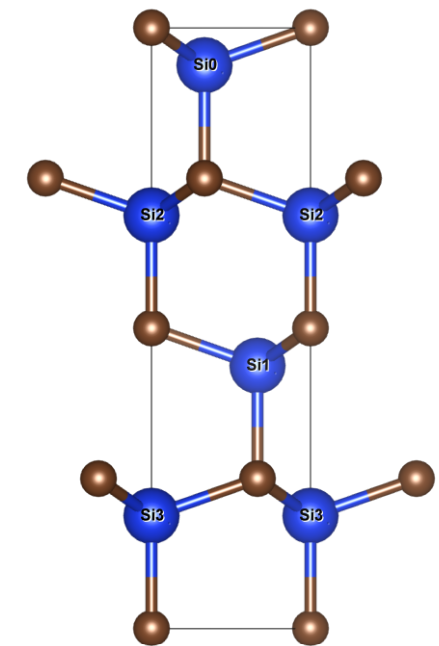}
\caption{4H-SiC crystal structure viewed from the c-axis (left) and a-axis (right). Hexagonal (h) sites are labeled as Si2 and Si3, while the quasi-cubic (k) sites are labeled as Si0 and Si1.}
\label{fig:ineq}
\end{figure}

The unit cell of 4H-SiC contains four silicon and four carbon atoms.
4H-SiC is a hexagonal polytype with a wurtzite structure characterized by a four-layer stacking sequence along the c-axis (\cref{fig:unit}). Its crystal structure is of the form ABCB, and it contains two inequivalent silicon sites and two inequivalent carbon sites, commonly referred to as hexagonal (h) and quasi-cubic (k) sites (\cref{fig:ineq}).

\subsection{Electronic Structure}

The electronic structure of pristine 4H-SiC is the reference against which defect-induced states are identified. The band structure and DOS calculations were performed using the PBE-GGA functional to provide reference values for the defect calculations and as verification of the calculation process.

\begin{figure}[ht]
\centering
\includegraphics[height=0.7\linewidth,width=0.85\linewidth]{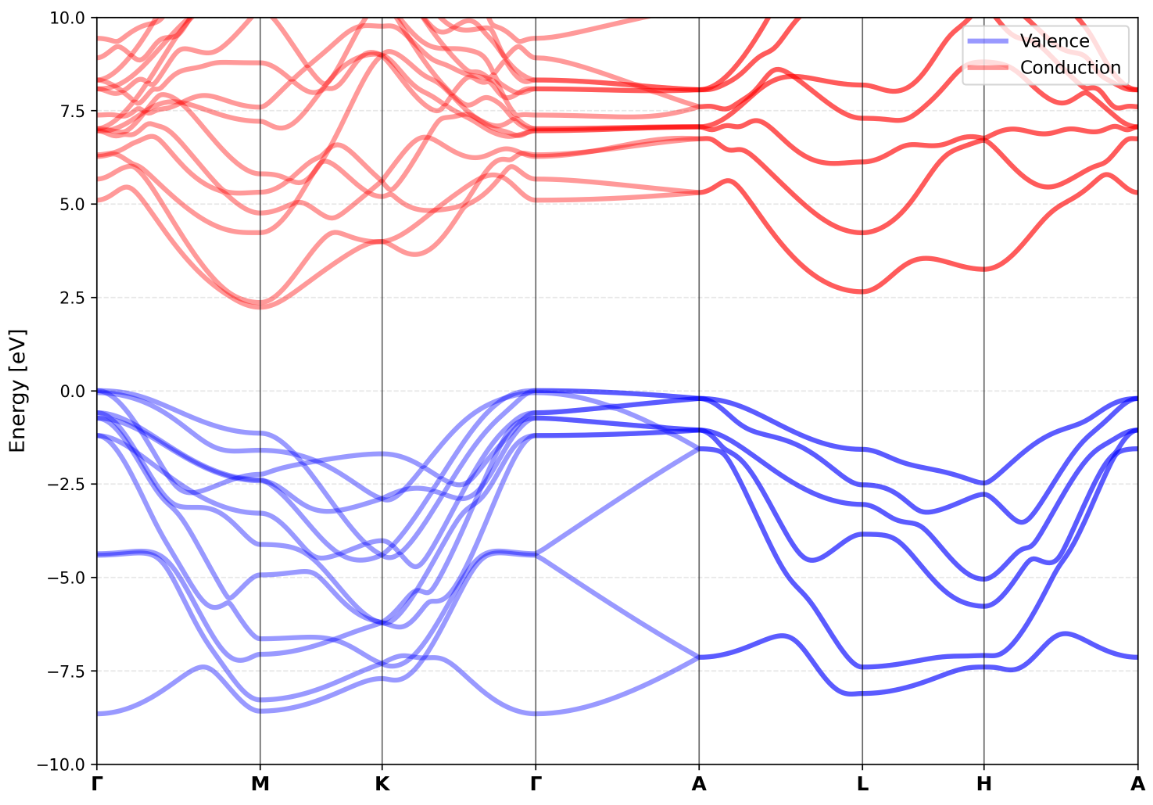}
\caption{Pristine 4H-SiC band structure.}
\label{fig:sicpbe}
\end{figure}

\begin{figure}[ht]
\centering
\includegraphics[height=0.85\linewidth,width=0.75\linewidth]{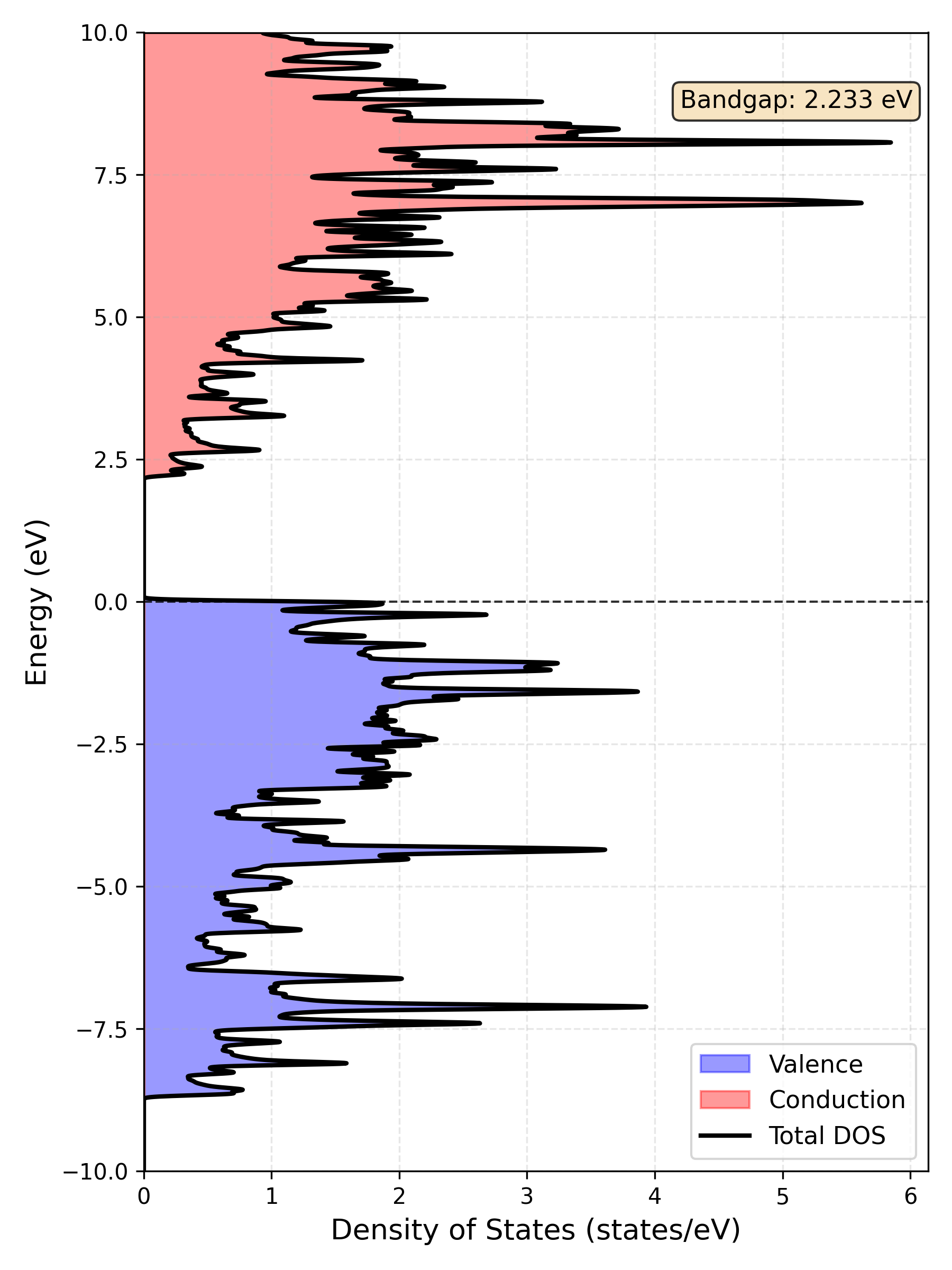}
\caption{Pristine 4H-SiC density of states.}
\label{fig:sicpbedos}
\end{figure}

As is typical for semilocal DFT, the PBE-GGA functional significantly underestimated the bandgap, yielding a calculated value of 2.23 eV, as shown in the pristine band structure (\cref{fig:sicpbe}) and in the DOS (\cref{fig:sicpbedos}). This is lower than the experimental bandgap of about 3.26 eV due to the self-interaction error inherent in GGA, but it aligns with other previous GGA calculations, validating the calculation method \cite{band1,band2,mat}.

%% file: sections/ErInSiC.tex
\section{Erbium point defects in SiC} \label{sec:ErInSiC}

In order to model a single point defect using DFT, the supercell expansion method was used, where the defect configuration was placed at the center of a 128-atom supercell of 4H-SiC. Due to the periodic assumptions of DFT, this corresponds to an erbium doping concentration of around 0.78$\%$. This has the potential to unintentionally modify the bulk band structure, a problem that can be mitigated with larger supercell volumes, as discussed in \cref{sec:conclusion}

\subsection{Defect Configurations}

After ion implantation and annealing, erbium ions can relax into a number of different defect configurations in 4H-SiC. For example, the erbium ion can settle into either of the two inequivalent silicon sites, and the surrounding lattice can have different levels of perturbation. This study considers four defect configurations:

\begin{itemize}
    \item  Substitutional erbium in an h-site (Er\textsubscript{h})
    \item  Substitutional erbium in a k-site (Er\textsubscript{k})
    \item Erbium-vacancy complex in an h-site (Er\textsubscript{h}V)
    \item Erbium-vacancy complex in a k-site (Er\textsubscript{k}V)
\end{itemize}

\begin{figure}[ht]
\centering
\includegraphics[height=0.7\linewidth,width=0.4\linewidth]{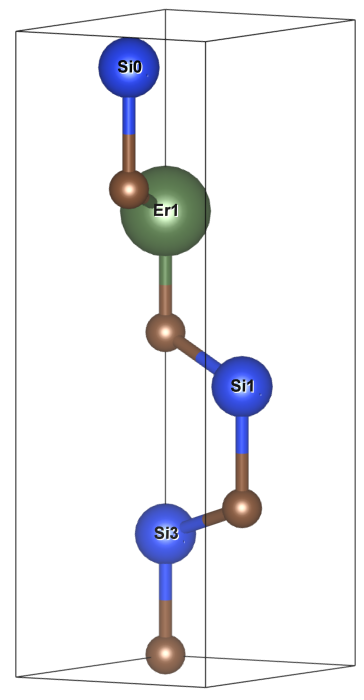}
\includegraphics[height=0.7\linewidth,width=0.4\linewidth]{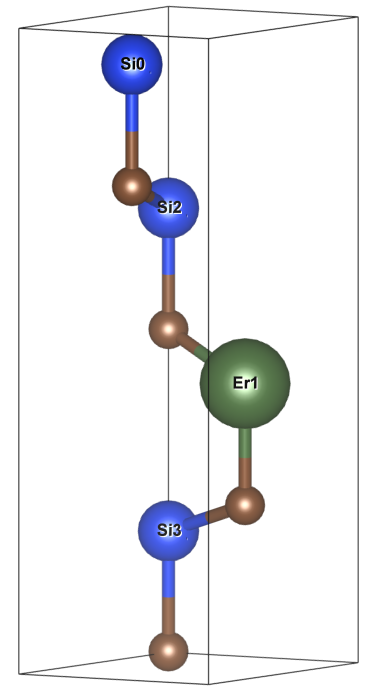}
\caption{Unit cell of 4H-SiC with an erbium ion replacing a silicon atom at an h-site (left) (Er\textsubscript{h}) and at a k-site (right) (Er\textsubscript{k}).}
\label{fig:com}
\end{figure}

In the substitutional configurations Er\textsubscript{h} and Er\textsubscript{k}, the surrounding lattice is unperturbed (\cref{fig:com}). In the erbium-vacancy complex configurations Er\textsubscript{h}V and Er\textsubscript{k}V, an adjacent carbon atom is missing in addition to the silicon atom that was replaced.

\subsection{Electronic Structure}

To identify defect-induced electronic states, the electronic band structure and DOS of four different supercells (one for each defect configuration) were calculated. Defect-induced states are characterized by their low dispersion due to the localized nature of the defects.

\begin{figure}[ht]
\centering
\includegraphics[height=0.8\linewidth,width=0.85\linewidth]{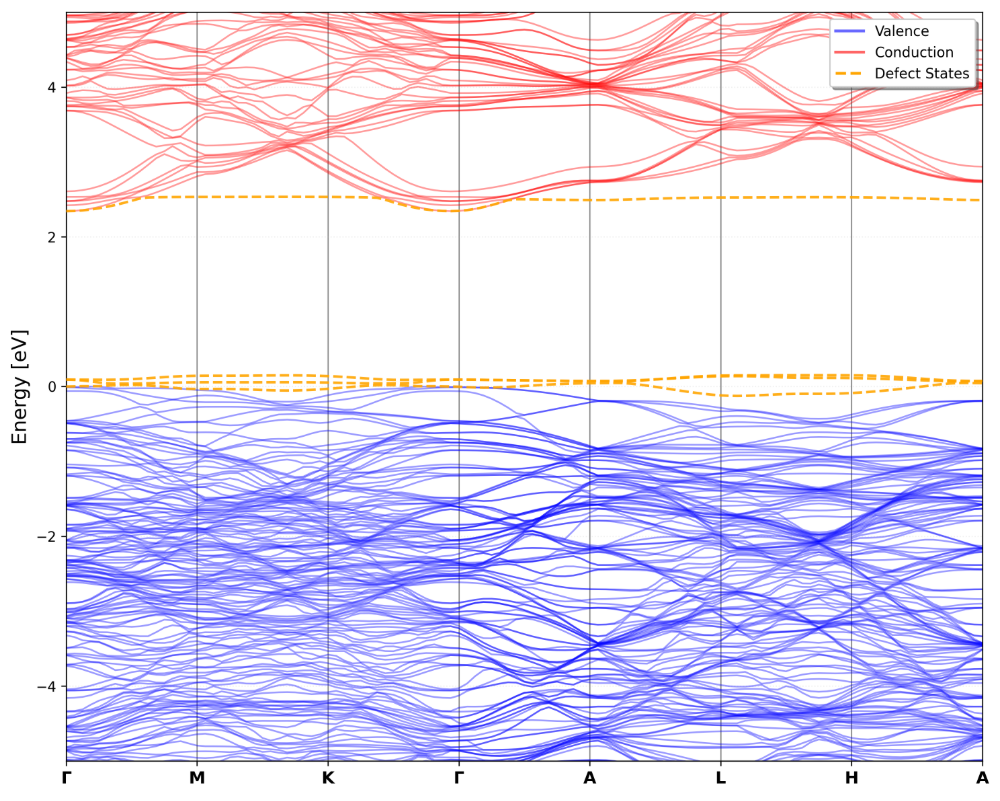}
\caption{Er\textsubscript{h} in 4H-SiC band structure.}
\label{fig:1}
\end{figure}

\begin{figure}[ht]
\centering
\includegraphics[height=0.85\linewidth,width=0.75\linewidth]{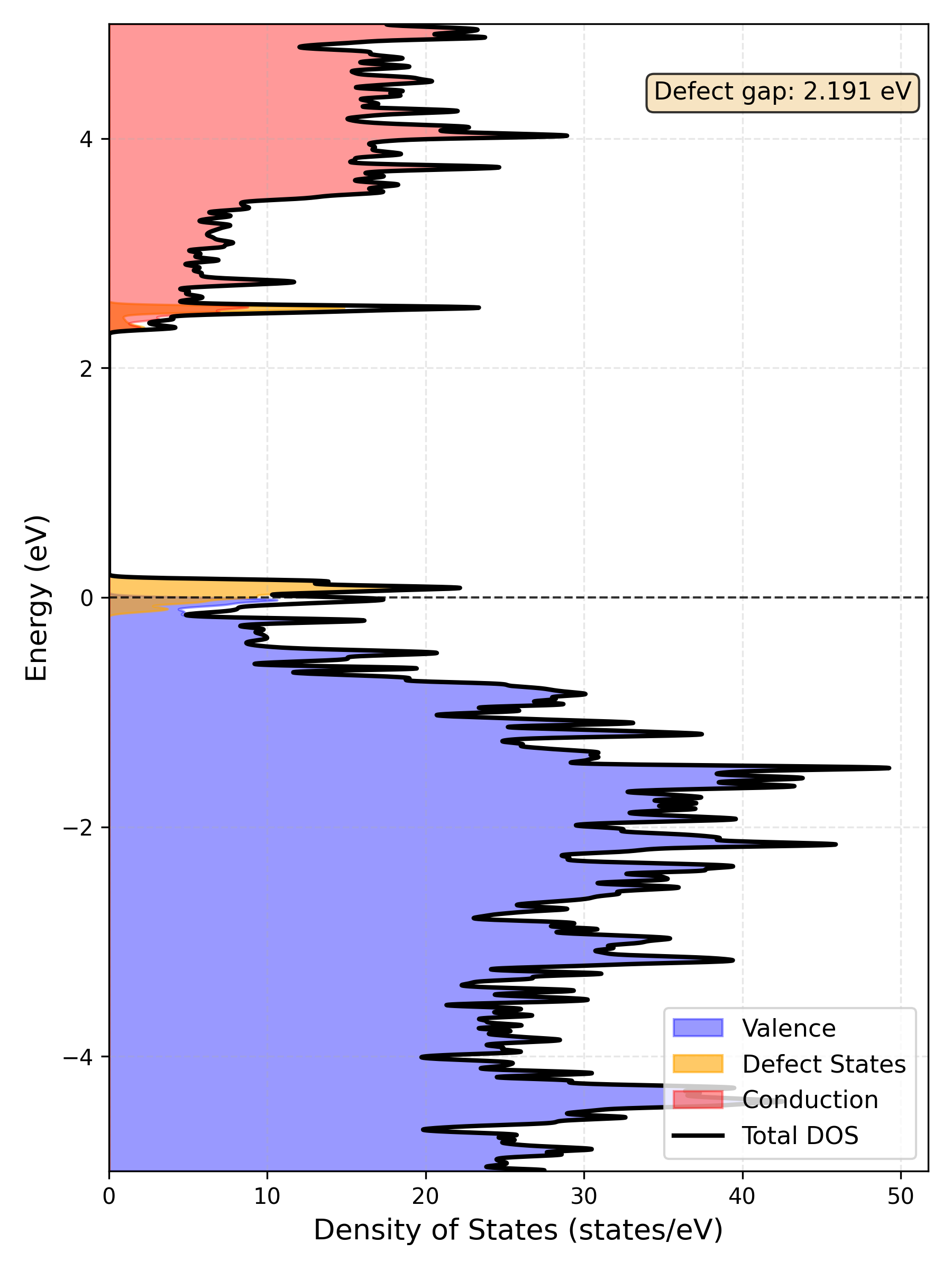}
\caption{Er\textsubscript{h} in 4H-SiC density of states.}
\label{fig:1d}
\end{figure}

The Er\textsubscript{h} calculated band structure (\cref{fig:1}) and corresponding DOS plot (\cref{fig:1d}) show defect-induced states both near the pristine conduction band minimum (CBM) and valance band maximum (VBM). The energy gap of the material between the defect-induced states was calculated to be around 2.19 eV.

\begin{figure}[ht]
\centering
\includegraphics[height=0.8\linewidth,width=0.85\linewidth]{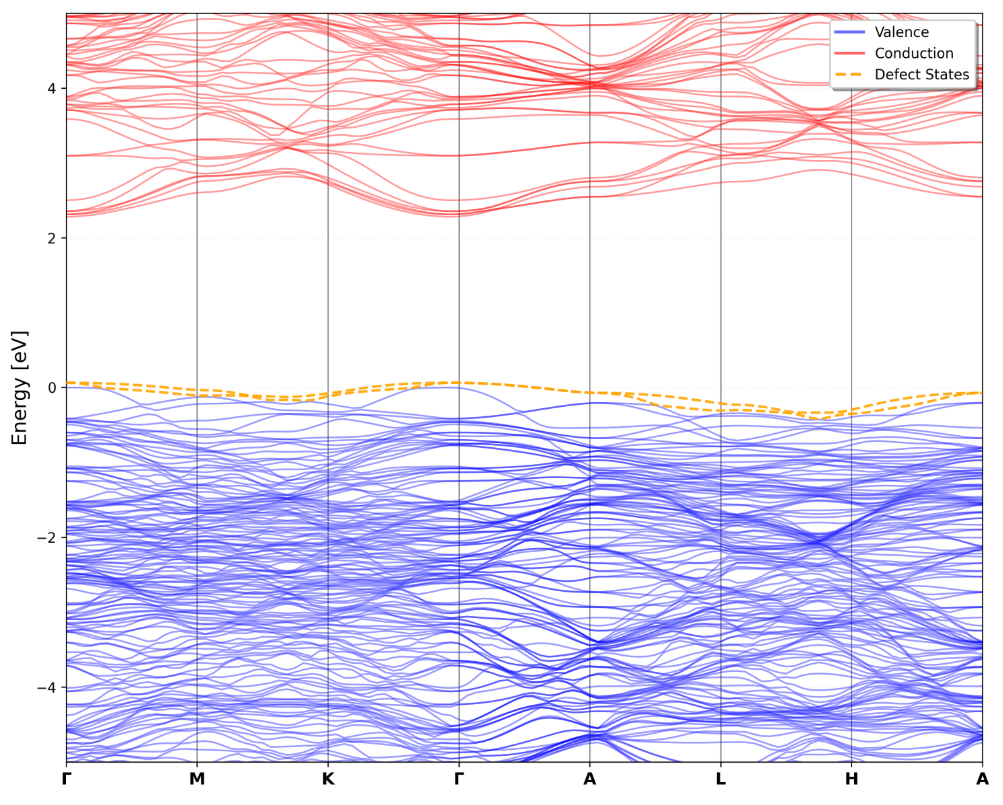}
\caption{Er\textsubscript{k} in 4H-SiC band structure.}
\label{fig:2}
\end{figure}

\begin{figure}[ht]
\centering
\includegraphics[height=0.85\linewidth,width=0.75\linewidth]{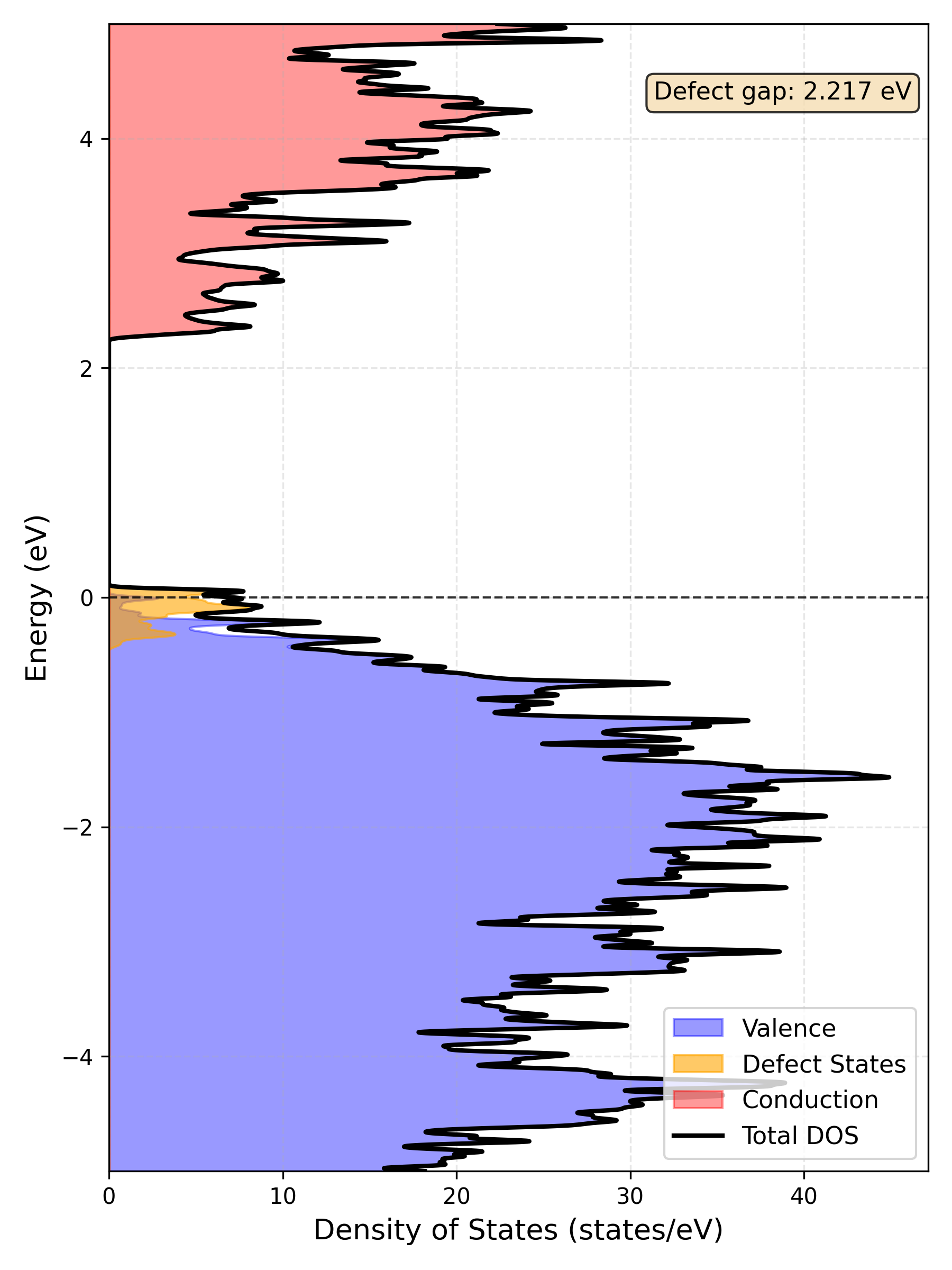}
\caption{Er\textsubscript{k} in 4H-SiC density of states.}
\label{fig:2d}
\end{figure}

Similar to the previous calculation, the Er\textsubscript{k} defect band structure (\cref{fig:2}) and DOS plot (\cref{fig:2d}) exhibit defect-induced states near the VBM, with an energy gap of around 2.22 eV. Unlike the previous calculation, they do not show any new states near the CBM. It is worth noting, however, that the SCF calculation for this configuration was unable to converge below the desired residual threshold after 100 iterations, and this is likely the reason for the lack of states in the bandgap near the CBM. This will be discussed in \cref{sec:conclusion}

\begin{figure}[ht]
\centering
\includegraphics[height=0.8\linewidth,width=0.85\linewidth]{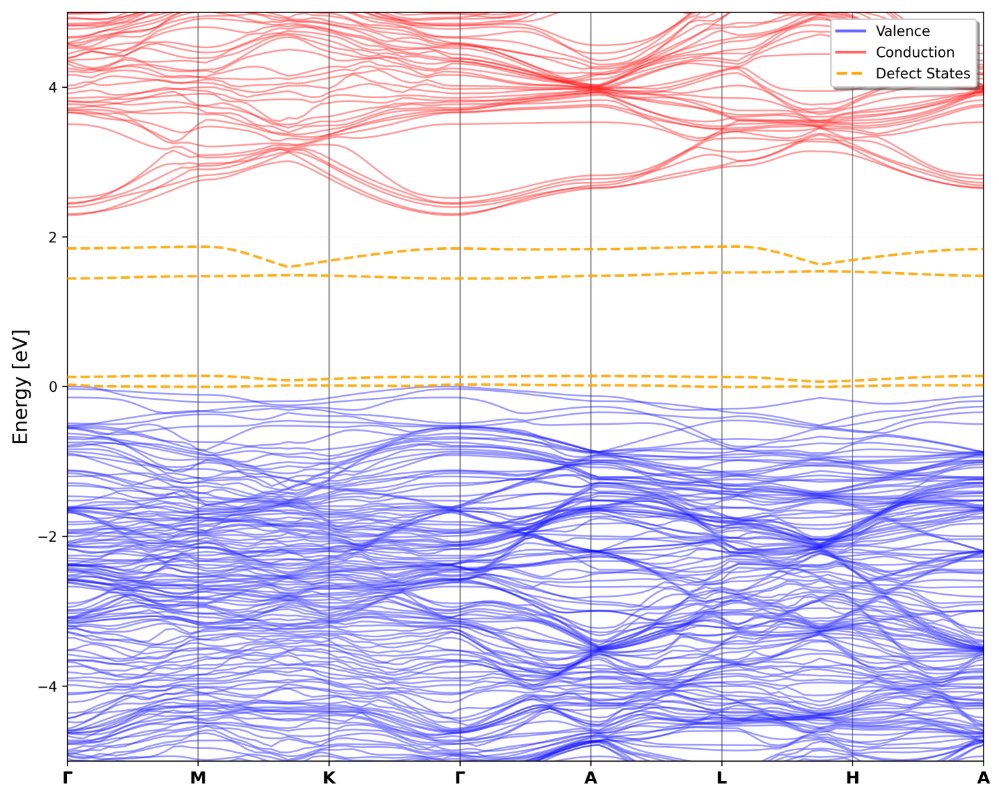}
\caption{Er\textsubscript{h}V in 4H-SiC band structure.}
\label{fig:3}
\end{figure}

\begin{figure}[ht]
\centering
\includegraphics[height=0.85\linewidth,width=0.75\linewidth]{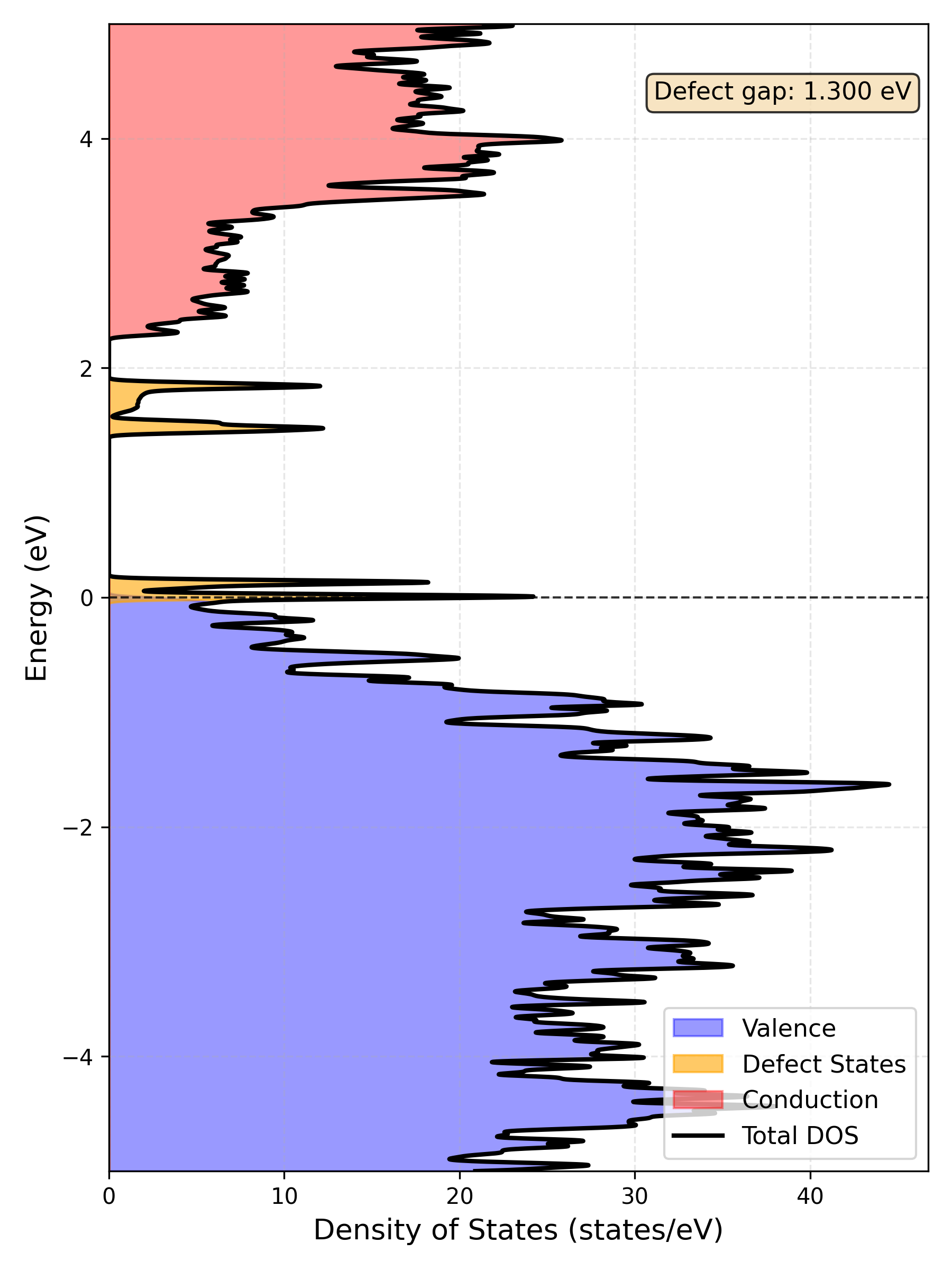}
\caption{Er\textsubscript{h}V in 4H-SiC density of states.}
\label{fig:3d}
\end{figure}

Unlike the pure substitutional configurations, the vacancy-complex configurations show well-defined defect states deep within the pristine bandgap. The Er\textsubscript{h}V band structure (\cref{fig:3}) and DOS (\cref{fig:3d}) show four distinct defect-induced energy states, with two near the pristine CBM and two near the pristine VBM. This configuration also resulted in a far-reduced energy gap between defect levels, around 1.3 eV.

\begin{figure}[ht]
\centering
\includegraphics[height=0.8\linewidth,width=0.85\linewidth]{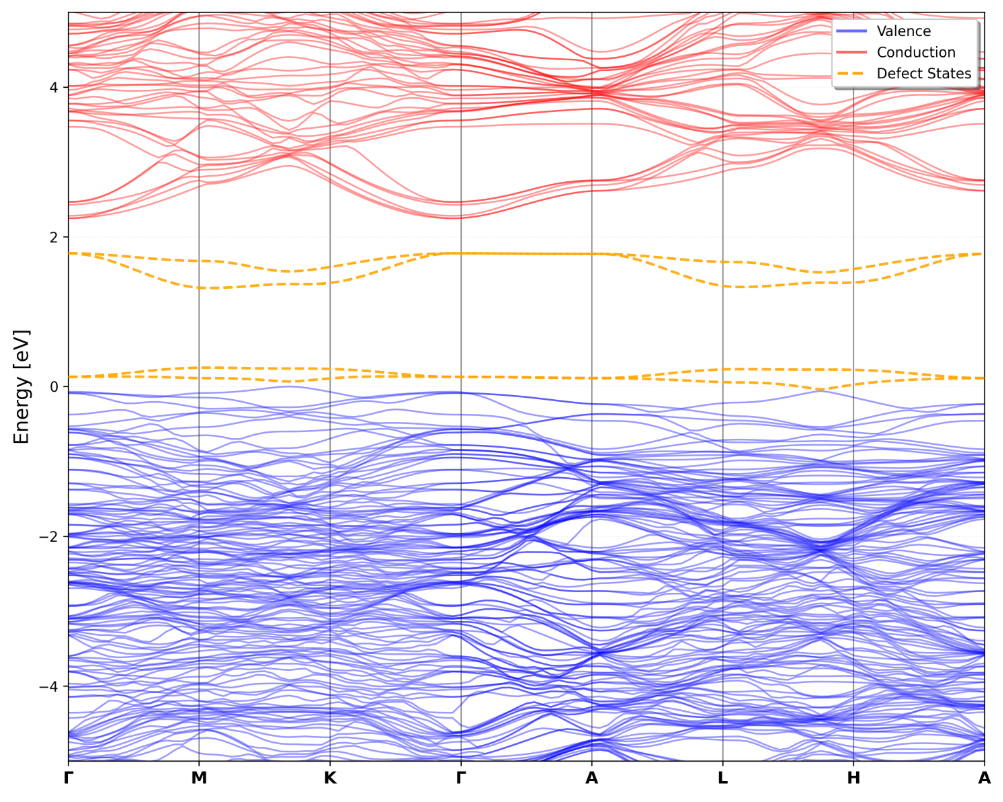}
\caption{Er\textsubscript{k}V in 4H-SiC band structure.}
\label{fig:4}
\end{figure}

\begin{figure}[ht]
\centering
\includegraphics[height=0.85\linewidth,width=0.75\linewidth]{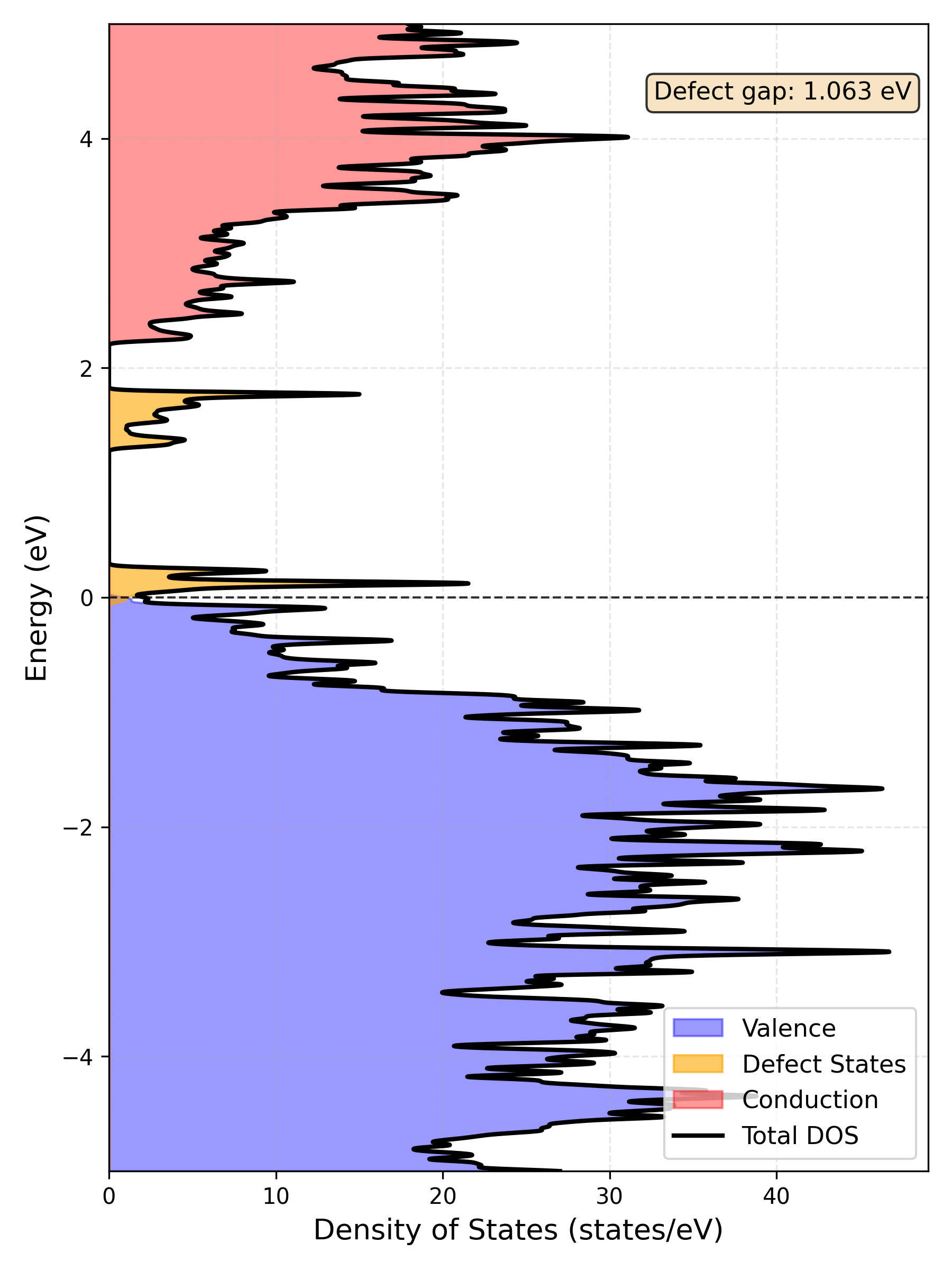}
\caption{Er\textsubscript{k}V in 4H-SiC density of states.}
\label{fig:4d}
\end{figure}

Much like the Er\textsubscript{h}V calculation, the Er\textsubscript{k}V defect configuration also displayed four defect-induced energy levels in its band structure (\cref{fig:4}) and DOS (\cref{fig:4d}), with two near the CBM and two near the VBM. These energy levels appeared deeper within the bandgap, leading to a new energy gap of around 1.06 eV.

Notably, none of the defect-induced energy gaps yielded the expected value of 0.8 eV from experimentation. This could be due to a number of factors, which will be discussed in \cref{sec:conclusion} Despite this, the presence of these states within the bandgap provides motivation for further investigation of this material in the future.

\subsection{Relative Formation Energy}

To assess the relative likelihood of each defect configuration forming during the implantation process and subsequent annealing, the relative formation energy of each configuration was calculated as 

\begin{center}
$E_R={E_T-E_{Def}}-E_0$.
\end{center}

In this equation, $E_R$ is the relative formation energy of each defect configuration, $E_T$ is the calculated total energy per atom for pristine 4H-SiC, $E_{Def}$ is the calculated total energy per atom for a given defect configuration, and $E_0$ is a chosen reference energy for normalization. Typically, formation energies are expressed in terms of total energies of the supercell with and without the defect, chemical potentials of the constituent species, and charge state corrections \cite{ErbiumPaper}. The simplified form used here compares the formation energies of the defect configurations relative to each other, which is sufficient for the purposes of this study.

The Er\textsubscript{k} configuration yielded the highest total energy during the SCF calculation, so this value was used as $E_0$. A lower relative formation energy indicates that the associated configuration is more stable than the reference and therefore indicates that it is more likely to form during implantation \cite{ErbiumPaper}.

The relative formation energies for each configuration (\cref{tab:form}) show that the Er\textsubscript{k} configuration has the highest $E_R$ by over 1 eV, indicating that it is an extremely unfavorable configuration relative to the others. However, this large value is almost certainly due to the lack of SCF convergence and is likely not an indication of the stability of the configuration itself.

\begin{table}[ht]
    \centering
    \begin{tabular}{l S} 
        \toprule
        \textbf{Defect Configuration} & \textbf{Relative Formation Energy ($E_R$)} \\
        \midrule
        Er\textsubscript{h} & -1.4815 \\
        Er\textsubscript{k} & 0.0000 \\
        Er\textsubscript{h}V & -1.3404 \\
        Er\textsubscript{k}V & -1.3402 \\
        \bottomrule
    \end{tabular}
    \caption{Relative formation energies (eV/atom) for each of the 4 defect configurations studied. Er\textsubscript{k} is used as the reference energy.}
    \label{tab:form}
\end{table}

The $E_R$ values of the other configurations are more informative. The most stable of the remaining configurations is Er\textsubscript{h}, with $E_R=-1.4815$ eV. This indicates that Er\textsubscript{h} is slightly more likely to form than either of the vacancy-complex configurations. Given the small difference (0.0002 eV) between the converged vacancy-complex formation energies, it is likely that the Er\textsubscript{k} formation energy when converged is near the converged value for Er\textsubscript{h}, but further investigation is required to confirm this.

%% file: sections/conclusion.tex
\section{Conclusion and Future Directions} \label{sec:conclusion}

This work presents a first-principles investigation of erbium-related point defects in 4H-SiC as a potential material platform for scalable quantum devices. Using density functional theory, the electronic structures and relative formation energies of four representative erbium defect configurations were analyzed, including substitutional erbium at both hexagonal and quasi-cubic silicon sites, as well as erbium-vacancy complexes at each site. The results demonstrate that erbium-related defects can introduce localized electronic states within the bandgap of 4H-SiC, supporting the
concept of defect-induced artificial atoms embedded within a wide-bandgap semiconductor host.

Among the configurations studied, the vacancy-complex defects Er\textsubscript{h}V and Er\textsubscript{k}V exhibited the most pronounced defect-induced states deep within the bandgap, resulting in significantly reduced effective energy gaps compared to both pristine 4H-SiC and pure substitutional defects. These localized states are desirable for quantum device applications, as they provide energetic isolation from the bulk bands and may support optically addressable transitions. In contrast, the pure substitutional configurations showed more limited defect state formation, particularly for the Er\textsubscript{k} configuration, likely due to the fact that it exhibited difficulties in achieving SCF convergence. Future studies should employ more robust convergence strategies, including increased iteration limits and improved initial charge density guesses, to achieve more reliable results for this configuration.

The relative formation energy analysis indicates that substitutional erbium at the hexagonal silicon site is the most stable of the converged configurations, suggesting that it may form preferentially during ion implantation and annealing. However, the negligible energy differences between the converged configurations, particularly between the vacancy-complex defects, imply that multiple erbium-related defect structures will likely coexist in experimentally realized samples.

Aside from the lack of SCF convergence for Er\textsubscript{k}, several limitations of the present study motivate important directions for future work. First, the supercell approach used here corresponds to an effective erbium concentration of 0.78\% due to periodic boundary conditions, which may artificially enhance defect-defect interactions and alter the overall band structure. Larger supercells should be employed to better approximate the isolated defect limit and to more accurately determine defect formation energies and electronic structure.

Second, the use of semilocal exchange-correlation functionals, even with the addition of a Hubbard U correction, is known to miscalculate bandgaps and can influence defect level positioning. This is one possible reason for the mismatch between the empirical defect-induced energy gap near 0.8 eV and the values obtained in these calculations. Future studies should incorporate hybrid functionals or many-body perturbation techniques to obtain more reliable absolute defect level energies.

Finally, spin-orbit coupling (SOC) was not explored in this work. This could potentially play a significant role in determining the stability and optical properties of this material due to the 4f electrons in erbium. Further calculations investigating DFT+U+SOC
would provide a more complete physical picture to compare to experimental results and may allow for more accurate calculations of optical transition energies between defect states.

In summary, this study establishes a first-principles foundation for erbium point defects in 4H-SiC as a viable platform for quantum photonic devices. The calculated electronic structure and relative stability of representative erbium defect configurations provide a theoretical basis for future experimental work. These findings demonstrate the potential for this platform's use in scalable quantum communication hardware.

%% file: sections/acknowledgments.tex
\newline
\newline
\begin{acknowledgments}

This work used the Bridges-2 system, which is supported by NSF award number OAC-1928147 at the Pittsburgh Supercomputing Center (PSC).

This material is based upon work supported by the National Science Foundation Graduate
Research Fellowship Program under Grant No. DGE2140739. Any opinions,
findings, and conclusions or recommendations expressed in this material are those of the
author(s) and do not necessarily reflect the views of the National Science Foundation.

\end{acknowledgments}

%% file: sections/bibliography.tex
\bibliographystyle{setup/bib_sophia}
\bibliography{references}